\documentclass[12pt,reqno]{amsart}

\usepackage{epsfig}
\usepackage{psfrag}
\usepackage{amsfonts}

\def\Tr{\mbox{Tr}\;}

\def\non{\nonumber}
\def\beq{\begin{equation}}
\def\eeq{\end{equation}}
\def\bit{\begin{itemize}}
\def\eit{\end{itemize}}
\def\beqn{\begin{eqnarray}}
\def\eeqn{\end{eqnarray}}
\newcommand{\ba}         {\begin{eqnarray}}
\newcommand{\ea}         {\end  {eqnarray}}
\newcommand{\ban}        {\begin{eqnarray*}}
\newcommand{\ean}        {\end  {eqnarray*}}

\newcommand{\la}[1]{\label{#1}}
\newcommand{\be}{\begin{equation}}
\newcommand{\ee}{\end{equation}}
\newcommand{\rmi}[1]{{\mbox{\scriptsize #1}}}
\newcommand{\rmii}[1]{{\mbox{\tiny\rm{#1}}}}
\newcommand{\fig}{Fig.~}
\newcommand{\eq}{Eq.~}
\newcommand{\se}{Sec.~}
\newcommand{\eqs}{Eqs.~}
\newcommand{\nr}[1]{(\ref{#1})}
\newcommand{\tr}{{\rm Tr\,}}

\newcommand{\fr}[2]{{\frac{#1}{#2}\,}}
\newcommand{\msbar}{{\overline{\mbox{\rm MS}}}}

\renewcommand{\vec}[1]{{\bf #1}}

\newcommand{\bmu}{\bar\mu}
\newcommand{\tinymsbar}{{\overline{\mbox{\tiny\rm{MS}}}}}

\newcommand{\Nf}{N_{\rm f}}
\newcommand{\Nc}{N_{\rm c}}

\newcommand{\rmO}{{\mathcal{O}}}
\newcommand{\CA}{\Nc}

\newcommand{\bG}{B_\rmi{G}}    
\newcommand{\tbG}{\tilde B_\tinymsbar}    
\newcommand{\mG}{m} 

\newcommand{\bE}[1]{\beta_\rmi{E#1}}
\def\lsi{\raise0.3ex\hbox{$<$\kern-0.75em\raise-1.1ex\hbox{$\sim$}}}
\def\gsi{\raise0.3ex\hbox{$>$\kern-0.75em\raise-1.1ex\hbox{$\sim$}}}
\newcommand{\lsim}{\mathop{\lsi}}
\newcommand{\gsim}{\mathop{\gsi}}
\newcommand{\e}{\epsilon}
\newcommand{\f}{\mbox{\sl f\,}}

\begin{document}

\begin{titlepage}
\begin{flushright}
     BI-TP 2006/13\\
     hep-ph/0605042
\end{flushright}
\par \vskip 15mm
\begin{center}
{\Large\bf 
  The leading non-perturbative coefficient in the \\[2mm] 
  weak-coupling expansion of hot QCD pressure
}
\end{center}
\par \vskip 10mm
\begin{center}
 {\bf F.\ Di Renzo}$\,^\rmi{a}$, 
 {\bf M.\ Laine}$\,^\rmi{b}$, 
 {\bf V.\ Miccio}$\,^\rmi{c}$, 
 {\bf Y.\ Schr\"{o}der}$\,^\rmi{b}$ and 
 {\bf C.\ Torrero}$\,^\rmi{b}$\\[5mm]
$^\rmi{a}${\sl Dipartimento di Fisica, Universit\`a di Parma, and} \\
~{\sl INFN, Gruppo Collegato di Parma, Parma, Italy}\\[.5 em]
$^\rmi{b}${\sl Faculty of Physics, University of Bielefeld, 
D-33501 Bielefeld, Germany} \\[.5 em]
$^\rmi{c}${\sl INFN, Sezione di Milano, Milano, Italy} \\[.5 em]
\vskip 2 mm
\end{center}
\par \vskip 5mm
\begin{center} { \large \bf Abstract} 
 \end{center}
\par \vskip 2mm
\begin{quote}
{
Using Numerical Stochastic Perturbation Theory
within three-dimensional pure SU(3) gauge theory, we estimate the last
unknown renormalization constant that is needed for converting the 
vacuum energy density of this model from lattice regularization to 
the $\msbar$ scheme. Making use of a previous non-perturbative lattice 
measurement of the plaquette expectation value in three dimensions,
this allows us to approximate the first non-perturbative coefficient 
that appears in the weak-coupling expansion of hot QCD pressure.

}
\end{quote}
\par \vskip 20mm
July 2006
\end{titlepage}

%
\section{Introduction}

It is well known that, despite asymptotic freedom, QCD can 
display non-perturbative phenomena even in situations where 
the system is characterised by a large momentum or mass 
scale $Q$, $Q \gg 1$~GeV. For instance, in deep inelastic 
scattering, power-suppressed contributions may arise, of the 
form $\sim E_\rmi{QCD}^n/Q^n$, where the numerator represents 
a non-perturbative contribution related to a certain 
quark or gluon condensate of dimension $n$,
arising from the Operator Product Expansion, and $E_\rmi{QCD}$ 
denotes the typical QCD energy scale, of a few hundred MeV. 

A conceptually similar situation arises when 
$Q$ is replaced by a high temperature $T$, and the observable
is replaced by minus the grand canonical free energy density, 
or the pressure, $p(T)$. The Operator Product Expansion gets 
then replaced by a construction of a low-energy effective field 
theory, which in the case of high temperatures amounts to 
dimensional reduction~\cite{dr}. The non-perturbative scale
$E_\rmi{QCD}$ gets replaced by that of the effective theory, 
$\sim \alpha_s T$~\cite{linde,gpy}, 
where $\alpha_s=g^2/4\pi$ is the strong coupling constant. In the effective
theory, the leading non-perturbative condensate has 
the dimension $n = 3$. Therefore, the formal weak-coupling
expansion of $p(T)$ contains non-perturbative coefficients,
starting at $\rmO(g^6)$~\cite{linde}.

While $\rmO(g^6)$ may appear to be an academically high order, recent advances
in perturbative QCD have made its determination an issue of 
practical importance~\cite{old,ma}. Indeed, perturbative corrections 
to the non-interacting Stefan-Boltzmann form of $p(T)$ 
have been determined at relative orders 
$\rmO(g^2)$~\cite{es}, 
$\rmO(g^3)$~\cite{jk}, 
$\rmO(g^4\ln(1/g))$~\cite{tt}, 
$\rmO(g^4)$~\cite{az}, 
$\rmO(g^5)$~\cite{zk,bn}, and 
$\rmO(g^6\ln(1/g))$~\cite{gsixg}, 
as a function of the number of colours, $\Nc$, the number
of massless quark flavours, $\Nf$, and, most recently, the 
chemical potentials, $\mu_i$, that can be 
assigned to the various quark flavours~\cite{av2}
(as long as they are small enough compared 
with the temperature, $|\mu_i| \lsim T/g$~\cite{muT}). 
Reaching the next unknown order $\rmO(g^6)$ depends, therefore, 
on the inclusion of the non-perturbative 
term.  Moreover, studies with high-temperature observables 
other than the pressure, in which analogous non-perturbative
coefficients arise but at lower orders, have shown 
that their inclusion is in general numerically important, 
in order to reliably determine the properties of hot QCD
at physically relevant temperatures~\cite{hp,gE2}.

By now, the first step has already been taken in order 
to determine the non-perturbative $\rmO(g^6)$ term: 
the gluon condensate of dimension $n=3$ was measured
with lattice Monte Carlo techniques 
within the dimensionally reduced effective field theory
in Ref.~\cite{plaq}.
Given that the other parts of the pressure computation have 
been formulated in the continuum $\msbar$ scheme, however, 
this result needs still to be converted 
to the same regularization~\cite{schemes,contlatt}.  
The purpose of the present paper is to finalise this task with
a certain numerical precision. 
Afterwards, only purely perturbative contributions 
in the $\msbar$ scheme remain to be determined,  
in order to complete the expression of $p(T)$ up to $\rmO(g^6)$
(in the notation of Ref.~\cite{gsixg}, $\bE{1}$ remains unknown, 
while $\bE{2}, \bE{3}$ have recently become available~\cite{bE2,gE2}).

We note that apart from formal interests,
the computations outlined above may also have phenomenological
relevance, in the contexts of cosmology and of heavy ion collision 
experiments. We do not elaborate on these issues in the present 
paper, however, since the current status on these fronts has very 
recently been summarised elsewhere~\cite{phen}.

The plan of this paper is the following. 
In \se~2 we set up the notation and describe
the overall strategy of the computation.
\se~3 contains a discussion of our computational tool, 
Numerical Stochastic Perturbation Theory in a covariant gauge.
\se~4 contains the data analysis and our results, 
while \se~5 draws some conclusions.

\vspace*{0.5cm}

%
\section{Basic definitions and overall strategy}

Let us start by considering three-dimensional (3d) pure SU($\Nc$)
gauge theory in dimensional regularization. 
The Euclidean continuum action can be written as 
\ba 
 S_\rmi{E} =  \int \! {\rm d}^d x\, \mathcal{L}_\rmi{E}
 \;, \qquad \la{eq:SE} 
 \mathcal{L}_\rmi{E} = \fr1{2g_3^2} \tr [F_{kl} F_{kl}]
 \;, \la{eq:LE}
\ea
where $d=3-2\epsilon$, 
$g_3^2$ is the (dimensionful) gauge coupling,  
$k,l=1,...,d$,  
$F_{kl} = i [D_k,D_l]$,  
$D_k = \partial_k + i A_k$, $A_k = A_k^BT^B$, 
$T^B$ are the Hermitean generators of SU($\Nc$), 
normalised as $\tr[T^A T^B] = \delta^{AB}/2$,
and repeated indices are assumed to be summed over.
Leaving out for brevity gauge fixing and Faddeev-Popov terms, 
the  ``vacuum energy density'' is defined as
\be
 \f_\tinymsbar \equiv - \lim_{V \to \infty} \frac{1}{V} 
 \ln \biggl[ \int \! \mathcal{D} A_k 
 \, \exp\Bigl( -S_\rmi{E} \Bigr) \biggr]_\tinymsbar
 \;, \la{eq:f}
\ee
where $V$ is the $d$-dimensional volume, 
$\mathcal{D} A_k$ a (gauge-invariant) functional integration measure, 
and we have assumed the use 
of the $\msbar$ dimensional regularization scheme to remove
any $1/\epsilon$ poles from the expression. 

The physical significance of $\f_\tinymsbar$ for high-temperature QCD
enters through the framework of dimensional reduction~\cite{dr,generic}.
Indeed, after the replacement 
$g_3^2 \to g^2 T[1 + \rmO(g^2)]$, $\f_\tinymsbar$ 
appears in the QCD pressure as an additive contribution, 
$\delta p(T) = - T \f_\tinymsbar$~\cite{linde,bn}.
The dependence on regularization (in particular, on the $\msbar$
scheme scale parameter $\bmu$) disappears, once all 
other contributions of the same order~\cite{gsixg} have been added.

Now, dimensional reasons and a perturbative computation of ultraviolet
divergences~\cite{sun} show that 
the structure of $\f_\tinymsbar$ is (after letting $\epsilon\to 0$)
\be
 \f_\tinymsbar = -g_3^6 
 \frac{d_A \CA^3}{(4\pi)^4} 
 \biggl[
  \biggl( \frac{43}{12} - \frac{157}{768} \pi^2 \biggr) 
 \ln\frac{\bmu}{2 \CA g_3^2} + \bG 
 \biggr]
 \;, \la{eq:structure}
\ee
where 
$d_A \equiv \Nc^2-1$.
The non-perturbative constant $\bG$,
~which actually 
is a function of $\Nc$, is what we would like to estimate
in the following. 
For future reference, we note that 
a logarithmic derivative of $\f_\tinymsbar$
with respect to $g_3^2$
immediately produces the gluon condensate: 
\ba
 \frac{1}{2 g_3^2} \Bigl\langle
 \tr [F_{kl} F_{kl}] 
 \Bigr\rangle_\tinymsbar 
 & = &  
 3 g_3^6 
 \frac{d_A \CA^3}{(4\pi)^4} 
 \biggl[
  \biggl( \frac{43}{12} - \frac{157}{768} \pi^2 \biggr) 
 \biggl( 
 \ln\frac{\bmu}{2 \CA g_3^2} -\fr13
 \biggr)
 + \bG 
 \biggr]
 \;.
 \la{eq:condensate}
\ea

We now go to lattice regularization. 
In standard Wilson discretization,  
the lattice action, $S_\rmi{L}$, corresponding to \eq\nr{eq:SE}, reads 
\ba
 S_\rmi{L} & = & 
 \beta \sum_{\bf x} \sum_{k < l}
 \Bigl[ 1 - \Pi_{kl}({\bf x})  \Bigr]
 \;, \la{eq:Sa}
\ea
where   
$
 \Pi_{kl}(\vec{x}) \equiv \mathop{\mbox{Re}} [ \Tr U_{kl}(\vec{x}) ]/\Nc 
$, 
$
 U_{kl}({\bf x}) \equiv U_k({\bf x})U_l({\bf x}+k)
 U_k^\dagger({\bf x}+l)U_l^\dagger({\bf x})
$
is the plaquette,
$U_k({\bf x})$ is a link matrix, 
${\bf x}+k\equiv  {\bf x}+a\hat\e_k$, where 
$a$ is the lattice spacing and $\hat\e_k$ 
is a unit vector,  and 
\be
 \beta \equiv \frac{2 \Nc}{g_3^2 a}
 \;. \la{eq:beta}
\ee
Note that the gauge coupling does not get renormalised in 3d, 
and the parameters $g_3^2$ appearing in \eqs\nr{eq:SE}, \nr{eq:beta} 
can hence be assumed finite and equivalent. 
The observable we consider is still the vacuum energy density, \eq\nr{eq:f}, 
which in lattice regularization reads
\be
 \f_\rmi{L} \equiv - \lim_{V \to \infty} \frac{1}{V} 
 \ln \biggl[ \int \! \mathcal{D} U_k 
 \, \exp\Bigl( -S_\rmi{L} \Bigr) \biggr]
 \;, \la{eq:fa}
\ee
where $\mathcal{D} U_k$ denotes integration over link matrices
with the gauge-invariant Haar measure.

Being in principle
physical quantities, 
the values of $\f_\tinymsbar$ and $\f_\rmi{L}$ must agree, 
provided that suitable vacuum counterterms are added to the theory.
Due to super-renormalizability, there can be such counterterms up to 4-loop
level only~\cite{framework}, and correspondingly
\ba
 \Delta \f & \equiv & 
 \f_\rmi{L} - 
 \f_\tinymsbar
 \nonumber \\ & = & 
 C_{1} \, \frac{1}{a^3} \biggl( \ln\frac{1}{a g_3^2} + C_1' \biggr )
 + 
 C_{2} \, \frac{g_3^2}{a^2} 
 + 
 C_{3} \, \frac{g_3^4}{a}
 + 
 C_{4} \, g_3^6 \biggl( \ln\frac{1}{a\bmu} + C_4' \biggr)
 + \rmO(g_3^8 a)
 \;, \la{eq:Deltaf}
\ea 
where the $C_i$ are dimensionless functions of $\Nc$. 
The values of $C_1,C_2,C_3,C_4$
are known, as we will recall presently; 
$C_1'$ is related to the precise normalisation
of the Haar integration measure and 
has no physical significance; and $C_4'$ 
will be estimated below. 

Correspondingly, the gluon condensates, \emph{i.e.}\ the 
logarithmic derivatives of $\f_\tinymsbar$, $\f_\rmi{L}$
with respect to $g_3^2$, 
can also be related by 
a perturbative 4-loop computation. Noting that three-dimensional rotational 
and translational symmetries 
allow us to write 
\be
 - g_3^2 \frac{\partial}{\partial g_3^2} 
 \f_\rmi{L}
 = \frac{3\beta}{a^3}
 \bigl\langle 1 - \Pi_{12} \bigr\rangle_\rmi{L}
 \;, \la{eq:plaq}
\ee  
and employing \eqs\nr{eq:condensate}, \nr{eq:Deltaf}, leads
to~\cite{plaq} 
\be
 8 \frac{d_A\CA^6}{(4\pi)^4} \bG =
 \lim_{\beta\to\infty} \beta^4
 \biggl\{
 \bigl\langle
 1 - \Pi_{12}
 \bigr\rangle_\rmi{L} - 
 \biggl[ 
 \frac{c_1}{\beta} + \frac{c_2}{\beta^2} + 
 \frac{c_3}{\beta^3} + \frac{c_4}{\beta^4}
 \Bigl(
 \ln\beta + c_4' 
 \Bigr)
 \biggr]
 \biggr\} 
 \;. \la{eq:betaG}
\ee
The values of the constants $c_1,...,c_4'$ are trivially
related to those of $C_1,...,C_4'$ in \eq\nr{eq:Deltaf}:
$c_1 = C_1/3$, $c_2 = - 2 \CA C_2/3$, $c_3 = -8 \CA^2 C_3/3$, 
$c_4 = - 8 \CA^3 C_4$ and, in particular, 
\be 
  c_4' = C_4' - \fr13 - 2 \ln(2\Nc)
  \;. \la{c4p} 
\ee
For $\Nc = 3$, the constants read~\cite{plaq,hk,PRyork,pt} 
\ba
 c_1 & = & \frac{d_A}{3} \approx 2.66666667
 \;, \la{eq:c1} \\[1mm]
 c_2 & = & 1.951315(2)
 \;, \la{eq:c2} \\[1mm]
 c_3 & = & 6.8612(2)
 \;, \la{eq:c3} \\[1mm]
 c_4 & = & 8 \frac{d_A\CA^6}{(4\pi)^4}
 \biggl(
 \frac{43}{12} - \frac{157}{768} \pi^2  
 \biggr)
 \approx 2.92942132
 \;. \la{eq:c4}
\ea
Moreover, lattice measurements~\cite{plaq} have shown that, for $\Nc = 3$, 
\ba
 \bG + 
 \biggl( \frac{43}{12} - \frac{157}{768} \pi^2 \biggr) c_4'
 = 10.7 \pm 0.4
 \;. \la{eq:final}
\ea

In order to extract $\bG$, as is our goal, 
we need to determine the unknown constant $c_4'$ in \eq\nr{eq:final}.
This can be achieved by  
repeating the same setup as above, but by regulating
infrared (IR) divergences through a mass regulator, $\mG$, 
instead of confinement. Indeed, the difference in \eq\nr{eq:Deltaf}
is IR insensitive, and does not change.
Thus we can extract $C_4'$ this way and, from \eq\nr{c4p}, $c_4'$. 
We denote quantities computed with a mass regulator with a tilde. 

With a mass regulator, 
the continuum computation produces~\cite{sun}
\be
 \tilde{\!\f}_\tinymsbar = ... -g_3^6 
 \frac{d_A \CA^3}{(4\pi)^4} 
 \biggl[
  \biggl( \frac{43}{12} - \frac{157}{768} \pi^2 \biggr) 
 \ln\frac{\bmu}{2 \mG} + \tbG(\alpha) 
 \biggr] + \rmO\Bigl(\frac{g_3^8}{\mG}\Bigr)
 \;, \la{eq:tildestructure}
\ee
where the lower order terms omitted vanish for $\mG\to 0$,
and $\tbG$ depends on the gauge parameter $\alpha$, since the 
introduction of a mass breaks gauge invariance.  
If, on the other hand,  
we carry out the same computation in lattice regularization, we expect
\be
 \tilde{\!\f}_\rmi{L} = ... - g_3^6 
 \frac{d_A \CA^3}{(4\pi)^4} 
 \biggl[
  \biggl( \frac{43}{12} - \frac{157}{768} \pi^2 \biggr) 
 \ln\frac{1}{a \mG} + \tilde B_\rmi{L}(\alpha) 
 + \rmO(\mG a)
 \biggr] + \rmO\Bigl(g_3^8 a,\frac{g_3^8}{\mG}\Bigr)
 \;, \la{tildefa}
\ee
where the lower order terms omitted go over to the ones
in \eq\nr{eq:Deltaf} for $\mG \to 0$. Given these forms
and the IR insensitivity of the difference in \eq\nr{eq:Deltaf}, 
we get
\ba
 \Delta \f & = & 
 \lim_{\mG \to 0} 
 \Bigl[\,\tilde{\!\f}_\rmi{L} - \tilde{\!\f}_\tinymsbar \Bigr] 
 \nonumber \\ & = & 
 ... 
 -  g_3^6 
 \frac{d_A \CA^3}{(4\pi)^4} 
 \biggl[
  \biggl( \frac{43}{12} - \frac{157}{768} \pi^2 \biggr) 
  \ln\frac{2}{a\bmu} + \tilde B_\rmi{L}(\alpha) - \tbG(\alpha)  
 \biggr]
 + \rmO(g_3^8 a)
 \;, 
\ea 
where the lower order terms omitted agree with \eq\nr{eq:Deltaf}.
Comparing the $\rmO(g_3^6)$ terms 
with \eq\nr{eq:Deltaf}, we can read off $C_4'$.
Subsequently, \eq\nr{c4p} gives $c_4'$.
Inserting finally the result from \eq\nr{eq:final}, we arrive at
the master relation
\be
 \bG = 10.7 \pm 0.4  - \tilde B_\rmi{L}(\alpha)  + \tbG(\alpha)
 + \biggl( \frac{43}{12} - \frac{157}{768} \pi^2 \biggr)
 \biggl(
  \fr13 + \ln2 + 2 \ln \Nc 
 \biggr)
 \;. \la{master}
\ee

It remains to determine $\tbG(\alpha)$ and $\tilde B_\rmi{L}(\alpha)$. 
Since the result of \eq\nr{master} is gauge independent, we 
choose the covariant Feynman gauge ($\alpha = 1$) in the following.  
A 4-loop continuum computation, described in Ref.~\cite{sun}, 
leads to a small set of fully massive master integrals that are known 
with high precision~\cite{sv}, and produces (see also Ref.~\cite{bk}) 
\be
 \tbG(1) =  
 -2.16562591949800919016...
 \;.
\ee
On the other hand, taking a logarithmic derivative with respect to 
$g_3^2$ from \eq\nr{tildefa}, we obtain, 
in complete analogy with \eq\nr{eq:betaG},  
\be
 8 \frac{d_A\CA^6}{(4\pi)^4} \tilde B_\rmi{L}(\alpha) =
 \lim_{\mG \to 0}\beta^4
 \biggl\{
 \left. \bigl\langle
 1 - \tilde \Pi_{12}
 \bigr\rangle_\rmi{L} \right|_\rmi{up to 4-loop}
 - 
 \biggl[
 \frac{c_1}{\beta} + \frac{c_2}{\beta^2} + 
 \frac{c_3}{\beta^3} + \frac{c_4}{\beta^4}
 \ln\frac{1}{a\mG} 
 \biggr]
 \biggr\} 
 \;. \la{eq:betaa}
\ee
Our task in the following is to compute the right-hand side
of this equation for $\alpha=1$ and, afterwards, to insert the 
result into \eq\nr{master}, in order to estimate $\bG$.

\pagebreak

%
\section{Numerical Stochastic Perturbation Theory}

To carry out the limit in \eq\nr{eq:betaa} requires a 4-loop
computation in lattice perturbation theory. As the master integrals
that appear in higher loop computations in lattice regularization need 
to be evaluated numerically in any case, we choose to carry out 
the whole computation numerically. This can be achieved through
the use of Numerical Stochastic Perturbation Theory (NSPT)~\cite{romm}, 
pioneered in recent years by the Parma group; 
a full account of the method can be found in Ref.~\cite{NSPT}. 

In its ``purest'' form, NSPT can be applied without either gauge 
fixing or masses as IR regulators. Since we compare with a dimensionally
regularized gauge fixed continuum computation with a mass as an 
IR regulator, however, we need to introduce the same tools in NSPT. 
The first two subsections describe our general implementation, 
and the third collects some technical details of the computation. 

%
\subsection{NSPT in a covariant gauge}

NSPT relies on Stochastic Quantization~\cite{ParWu}
(for an extensive review, see Ref.~\cite{rev}). 
In this approach quantum fields are given an extra coordinate, $\tau$, 
which is to be regarded as a stochastic time in which an evolution takes 
place according to the Langevin equation. This is in close analogy
with the ``time'' evolution of the Markov chain that is used in standard
Monte Carlo simulations; indeed, Stochastic Quantization can also 
be used for Monte Carlo simulations~\cite{Batrouni}
(for a concise review, see Ref.~\cite{ak}).

For lattice gauge theories, the Langevin equation reads 
\begin{equation}
  \partial_\tau U_{k,\eta}(\vec{x},\tau) =
  -i\Bigl\{ \nabla_{k,\vec{x}}\, S[U_{k,\eta}]+\eta_{k}(\vec{x},\tau)
  \Bigr\} \; U_{k,\eta}(\vec{x},\tau) \;, \label{LangEq}
\end{equation}
where we assume the use of lattice units ($a=1$) in the 
spatial directions. The derivative 
$\nabla_{k,\vec{x}}$ is defined~\cite{ddh} as
$\nabla_{k,\vec{x}} \equiv T^A \nabla_{k,\vec{x}}^A$, 
with
$
 \nabla_{k,\vec{x}}^A S[U_k(\vec{x})] \equiv \lim_{\epsilon\to 0}
 \{S[\exp(i \epsilon T^A) U_k(\vec{x})] - S[U_k(\vec{x})] \}/\epsilon 
$, 
where $T^A$ are 
the generators in the fundamental representation, normalised as before.
Moreover, $\eta_k$ is a gaussian noise in the adjoint representation, 
$\eta_k \equiv \eta_k^A T^A $. 
In Eq.~(\ref{LangEq}) we adhere to a precise notation 
in which the dependence of the solution on the stochastic 
noise is explicitely shown. Since this notation is a bit pedantic, 
we will drop it in the following. It is worth 
stressing that the evolution dictated by \eq\nr{LangEq} preserves 
unitarity, and that $\nabla_{k,\vec{x}}$
is consistent with partial integration over the Haar measure. 

The main assertion of Stochastic Quantization is that the path integral 
correlation functions of the field theory, computed with the Haar measure, 
can be traded for stochastic time averages
in the asymptotic $\tau \rightarrow \infty$ limit: 
\be
 \frac{1}{Z} \int \! \mathcal{D} U_k \, 
 \mathcal{O}[U_k(\vec{x})]\exp({-S}) 
 = \lim_{\tau\to\infty} \frac{1}{\tau}
 \int_0^\tau \! {\rm d} \tau' \, 
 \mathcal{O}[U_k(\vec{x},\tau')]
 \;, 
\ee
where $\mathcal{O}$ is some observable.  
NSPT is obtained by expressing the 
solution of the Langevin equation as a power series in the coupling 
constant and by numerically integrating the hierarchy of equations 
that results from inserting this expansion into Eq.~(\ref{LangEq}). 
In our notation the expansion reads~\cite{romm,NSPT} 
\begin{equation}
  U_k(\vec{x}) = 
  1 + \sum_{i=1}^{N} \, \beta^{-\frac{i}{2}} \; U_k^{(i)}(\vec{x}) 
  \;, \la{link_expansion}
\end{equation}
in which $N$ is the highest order one wants to reach in the computation. 
In our case we need to expand the field up to $\beta^{-4}$, that is $N=8$. 
Note that since \eq\nr{LangEq} guarantees unitarity by 
construction, the terms $U_k^{(i)}(\vec{x})$ will automatically 
inherit the corresponding properties.


While the spirit of Parisi's and Wu's original paper~\cite{ParWu} was
to offer a possibility for performing perturbative computations without 
gauge fixing, 
gauge fixing can naturally be added to the framework through 
the Faddeev-Popov mechanism, like to lattice gauge theory in general.  
The partition function is written as
\begin{equation}
   Z = \int\! \mathcal{D} U_k \,
  {\det \Delta_\rmi{FP}}
  \exp\big(\!-S_\rmi{L}\!-S_\rmi{GF}\big)
  \;, \la{FP_Z}
\end{equation}
where $S_\rmi{L}$ is the Wilson action in \eq\nr{eq:Sa}. 
For $S_\rmi{GF}$, we choose the standard covariant form, 
\begin{equation}
    S_\rmi{GF} =  \frac{\beta}{4\Nc\alpha}\sum_{\vec{x},~\!\!A}
 \Bigl[ 
 \hat \partial_{k}^{L}\phi_{k}^{A}(\vec{x})\Bigr]^2
 \;, \la{GF}
\end{equation}
where $\alpha$ is the gauge parameter;
$\hat \partial_{k}^{L}$ is the left difference operator;
we have defined Lie algebra valued fields $\phi_k$
through $U_k = \exp(i \phi_k)$;
and we have written $\phi_k = \phi_k^A T^A$. 
The naive continuum limit is obtained through
the identification $\phi_k = a g_3 A_k$.
Once we enforce the expansion in \eq\nr{link_expansion} for the 
links, the Lie algebra valued fields are expanded as well: one simply 
needs the Taylor series for $\phi_k = -i \ln[1 + \delta U_k]$. 

Note that in standard lattice perturbation theory 
\eq\nr{FP_Z} is further modified by writing the 
functional integration in terms of $\phi_k$. 
On the contrary, we directly expand in 
terms of the link variables $U_k$ (cf.\ \eq\nr{link_expansion}), 
in terms of which Eq.~(\ref{LangEq}) is formulated. Therefore we do not 
need to add any ``measure term'' to the action. Nevertheless, 
the values of various operator expectation values remain identical,
order by order in $\beta^{-1}$, to those in the standard
perturbative framework.
 
The Faddeev-Popov operator corresponding to
the gauge function in \eq\nr{GF} reads
$
 \Delta_\rmi{FP} \equiv
  -\hat \partial_{k}^{L}
   \hat{D}_{k}[\phi]
$.
Since $S_\rmi{GF}$ is expressed in terms of Lie
algebraic fields, $\hat{D}_{k}[\phi]$ comes from evaluating the
response of the field $\phi_k$ to a gauge transformation. The latter 
is defined in terms of the link variables $U_k$, so it
does not come as a surprise that $\hat{D}_{k}[\phi]$ is not
expressed in closed form. It can however be written as a perturbative 
expansion, whose first terms read~(see e.g. Ref.~\cite{Rothe}, 
whose notations we follow) 
\be
 \hat{D}_{k}[\phi](\vec{x}) = 
 \biggl[1+\frac{i}{2}\displaystyle \Phi_{k}(\vec{x})
 -\frac{1}{12} \displaystyle \Phi_{k}^2(\vec{x})
 -\frac{1}{720} \displaystyle \Phi_{k}^4(\vec{x})
 -\frac{1}{30240}\displaystyle \Phi_{k}^6(\vec{x})
  +\rmO\Big(\displaystyle \Phi_{k}^8\Big)\biggr]
  \, \hat \partial_{k}^{R}+i\displaystyle \Phi_{k}(\vec{x})
 \;,
\ee
where $\hat\partial_{k}^{R}$ is the right difference operator, 
while $\Phi_k$ is the field in the adjoint representation,
\begin{equation}
        \displaystyle \Phi_{k}(\vec{x}) \equiv
        \phi_{k}^A(\vec{x})~\!F^A\;, 
        \;\;\;\;\;\; (F^A)_{BC}=-if^{ABC}
        \;.
\end{equation}
In order to set up the proper Langevin equation, 
we finally rephrase the Faddeev-Popov determinant 
as a new contribution to the action, 
$\det \Delta_\rmi{FP} = \exp(\tr\ln \Delta_\rmi{FP})$. 

\subsection{Mass regulator}
We still need to add a mass regulator to the gauge-fixed framework. 
We note that the same mass is given to the gluon and the ghost 
fields in the continuum computation we want to match to. We
therefore modify the total action to become
\begin{equation}
        \tilde S \equiv \tilde S_\rmi{L} + S_\rmi{GF} + \tilde S_\rmi{FP}
        \;.
\end{equation}
The gluonic action has been modified by a mass term, 
\begin{equation}
   \tilde S_\rmi{L} \equiv S_\rmi{L} + 
   \frac{\beta~\!(a \mG)^2}{4\Nc}\!\! 
   \sum_{\vec{x}}\!\phi_{k}^{A}(\vec{x})\phi_{k}^{A}(\vec{x})
   \;,
\end{equation}
and the Faddeev-Popov action reads
\begin{equation}
   \tilde S_\rmi{FP} \equiv - \tr \ln \tilde \Delta_\rmi{FP}
   \;, \quad \tilde \Delta_\rmi{FP} \equiv - 
   \hat \partial_{k}^{L}\hat{D}_{k}[\phi]+
   (a \mG)^2
   \;. \la{tildeL}
\end{equation}
Even if we never make use of ghost fields in our approach, 
\eq\nr{tildeL} amounts to giving them a mass, 
as is clear from writing
the operator $\tilde \Delta_\rmi{FP}$ in Fourier-representation.

%
\subsection{Some technical implementation issues}

To treat the Faddeev-Popov determinant as a part of the action 
means that, because of the Langevin equation, one has to face the 
quantity 
$
  \nabla_{k,\vec{x}} \tilde S_\rmi{FP} = 
    - \nabla_{k,\vec{x}} \tr[\ln \tilde \Delta_\rmi{FP}]
  = - T^A \tr[\nabla_{k,\vec{x}}^A \tilde \Delta_\rmi{FP} \; 
  \tilde \Delta_\rmi{FP}^{-1}]
$, 
with $\tilde \Delta_\rmi{FP}$ from \eq\nr{tildeL}. We follow the
procedure proposed in Ref.~\cite{Lat98} (see also Ref.~\cite{Batrouni}). 
Our study is actually the first practical implementation, 
but the essential ingredients 
are the same as for the treatment of the fermionic determinant 
in unquenched NSPT, as discussed in Ref.~\cite{NSPT}.

By introducing an extra gaussian noise $\xi$ 
normalized as $\langle \xi^*_k \xi_n \rangle_{\xi} = \delta_{kn}$, 
one can substitute\footnote{%
 Here ${k,l,n}$ should be regarded as multi-indices, 
 including space coordinates and colour.
 }
\begin{equation}
\label{introXI}
        - \tr[\nabla_{k,\vec{x}}^A \tilde \Delta_\rmi{FP} \; 
          \tilde \Delta_\rmi{FP}^{-1}] 
     \longrightarrow 
        - \langle \mathop{\mbox{Re}} \{\xi_k^*(\nabla_{k,\vec{x}}^A 
        \tilde \Delta_\rmi{FP})_{kl}
      (\tilde \Delta_\rmi{FP}^{-1})_{ln}\xi_n \} \rangle_{\xi} \;.
\end{equation}
Taking the real part would be unnecessary 
after the average $\langle ... \rangle_\xi$, 
but we find it convenient to impose it already before the averaging.
The advantage of the form introduced is made clear by rewriting
\begin{equation}
\label{XIandPSI}
       \xi_k^*(\nabla_{k,\vec{x}}^A \tilde \Delta_\rmi{FP})_{kl}
      (\tilde \Delta_\rmi{FP}^{-1})_{ln}\xi_n 
       \equiv 
        \xi_k^*(\nabla_{k,\vec{x}}^A 
        \tilde \Delta_\rmi{FP})_{kl}\psi_l 
       \;,
\end{equation}
where $\psi \equiv \tilde \Delta_\rmi{FP}^{-1}\,\xi$.
In NSPT, then,  we need to compute
\begin{equation}
\label{psidef}
        \psi^{(i)} \equiv (\tilde \Delta_\rmi{FP}^{-1})^{(i)}   
        \; \xi \;.
\end{equation}
It is worth stressing that the noise $\xi$ has no power expansion, 
while the field $\psi$ 
(like any other field in NSPT) is expanded because 
of the power expansion of $\tilde \Delta_\rmi{FP}^{-1}$ (which 
is a function of the fields $\phi$, \emph{i.e.}\ of the fields $U$). 

That \eq\nr{psidef} can be evaluated efficiently  
within the NSPT framework stems from the expansion of the 
operator $\tilde \Delta_\rmi{FP}^{-1}$. 
Once a generic matrix $M$ is given as a power expansion,
\be
  M = M^{(0)} + \sum_{i=1}^{\infty} \beta^{-\frac{i}{2}} M^{(i)} \;,
\ee
the expansion for its inverse reads
\begin{eqnarray}
        M^{-1}  
       &=& {[M^{(0)}]}^{-1} 
      + \sum_{i=1}^\infty  \beta^{-\frac{i}{2}} {[M^{-1}]}^{(i)}
      \;. 
\end{eqnarray}
Here the non-trivial terms are obtained through 
a simple recursive relation: 
\begin{eqnarray}\label{eq:rec}
{[M^{-1}]}^{(1)} &=& - {[M^{(0)}]}^{-1} M^{(1)} {[M^{(0)}]}^{-1} \;, \non \\
{[M^{-1}]}^{(2)} &=& - {[M^{(0)}]}^{-1} M^{(2)} {[M^{(0)}]}^{-1} 
               - {[M^{(0)}]}^{-1} M^{(1)} {[M^{-1}]}^{(1)} \;, \non \\
\ldots          && \non \\
{[M^{-1}]}^{(i)} &=& - {[M^{(0)}]}^{-1} \; \sum_{j=0}^{i-1} 
  \, M^{(i-j)} {[M^{-1}]}^{(j)} \;.
\end{eqnarray}
In our case this leads to 
\begin{eqnarray}\label{eq:xij}
 \psi^{(0)} &=& {[\tilde \Delta_\rmi{FP}^{(0)}]}^{-1} \xi \;, \non \\
 \psi^{(1)} &=& - {[\tilde \Delta_\rmi{FP}^{(0)}]}^{-1} 
    \tilde \Delta_\rmi{FP}^{(1)} \psi^{(0)} \;, \non \\
 \psi^{(2)} &=& - {[\tilde \Delta_\rmi{FP}^{(0)}]}^{-1} 
   \left[\tilde \Delta_\rmi{FP}^{(2)} \psi^{(0)} + 
 \tilde \Delta_\rmi{FP}^{(1)} \psi^{(1)} \right] \;, \non \\
 \ldots           \non \\
 \psi^{(i)} &=& - {[\tilde \Delta_\rmi{FP}^{(0)}]}^{-1} 
  \; \sum_{j=0}^{i-1} \, \tilde \Delta_\rmi{FP}^{(i-j)} \psi^{(j)} \;.
\end{eqnarray}
Eq.~(\ref{eq:xij}) states that there is no actual 
matrix inversion to take. Indeed, 
${[\tilde \Delta_\rmi{FP}^{(0)}]}^{-1}$ 
is independent of the $\phi$ fields and 
its expression is well known: it is 
the ({\em would-be}) ghost propagator, 
diagonal in Fourier space. 
Note that the mass regulator makes it well-defined 
at every value of the momentum. 
As for the various orders $\psi^{(i)}$, 
they are naturally computed by iteration. 
At every order only one application 
of ${[\tilde \Delta_\rmi{FP}^{(0)}]}^{-1}$ 
is needed: this propagator operates on a sum of 
already computed quantities (the lower order $\psi$'s). 
While ${[\tilde \Delta_\rmi{FP}^{(0)}]}^{-1}$ is diagonal in 
momentum space, all the other operators are almost diagonal 
in configuration space. This suggests the 
strategy of going back and forth from 
Fourier space via a Fast Fourier Transform. 
It remains to point out that also the expression 
for $\nabla_{k,\vec{x}}^A \tilde \Delta_\rmi{FP}$ 
(and its power expansion) is substantially local, 
so that the big 
inner product in Eq.~(\ref{XIandPSI}) 
is not too difficult to deal with. 
Finally, $\tilde S_\rmi{FP}$ and 
$\tilde \Delta_\rmi{FP}$ are naturally 
written in the adjoint representation, so that one 
has to devise an efficient way of dealing with 
cascades of commutators of the $\phi$ fields.

In order to solve \eq\nr{LangEq} numerically, the stochastic time
variable $\tau$ needs to be discretised: $\tau \equiv n a_\tau$, where
$n$ is an integer. We use
different values of $a_\tau$, average over each thermalised signal, 
and then extrapolate in order to get the value of the 
desired observable at $a_\tau = 0$. 
\eq\nr{LangEq} is discretized in the standard way~\cite{Batrouni}
which automatically preserves the unitarity 
of our degrees of freedom:
\begin{equation}\label{iteration}
 U_{k}(\vec{x},(n+1) a_\tau) = 
 e^{- i F_{k}(\vec{x},n a_\tau)[U,\eta]} \,\, 
 U_{k}(\vec{x},n a_\tau)
 \;,
\end{equation}
where
\begin{equation}
  F_{k}(\vec{x},n a_\tau)[U,\eta] \, = 
  a_\tau \nabla_{k,\vec{x}} \tilde S[U] + 
  \sqrt{a_\tau} \, \eta_{k}(\vec{x},n a_\tau) \;,
\end{equation}
and we have assumed the normalization
$
 \langle \eta(\vec{x},m a_\tau)\eta(\vec{y}, n a_\tau) \rangle_\eta
 = 2 \delta_{\vec{x}\vec{y}} \delta_{mn}
$.

We note that \eq\nr{iteration} is 
only accurate to first order in $a_\tau$.
As a consequence, if the action is written as a sum 
($\tilde S \equiv \tilde S_\rmi{L} + S_\rmi{GF} + \tilde S_\rmi{FP}$), 
one can to the same accuracy rewrite \eq\nr{iteration} as 
\begin{equation}
 U_{k}(\vec{x},(n+1) a_\tau) = e^{- i F^{(2)}_{k}[U]} \,\, 
 \left\{ \,\,e^{- i F^{(1)}_{k}[U,\eta]} \,\, 
 U_{k}(\vec{x},n a_\tau) \,\, \right\}
 \;. 
\end{equation}
The advantage of this form is that 
from the point of view of program implementation, 
it is easier to first evolve the field by 
\be
 F^{(1)}[U,\eta] = \, a_\tau \nabla_{k,\vec{x}} S_\rmi{L} 
  + \sqrt{a_\tau} \, \eta_k
\ee
(\emph{i.e.}\ the contribution coming from 
standard Wilson action plus gaussian noise) and then 
by
\be
 F^{(2)}[U] = \, a_\tau \nabla_{k,\vec{x}} (\tilde S - S_\rmi{L}) \; .
\ee
Indeed, $F^{[2]}[U]$ depends on the $U$'s 
only through the $\phi$'s, \emph{i.e.}\ 
the first step is performed in terms of the $U$ fields, 
the second in terms of the  $\phi$ fields. 
It is also easy to realize that the first step can be 
implemented as a sequential sweep through the lattice, while the second one 
requires the construction of a \emph{global} contribution 
(the inner product in Eq.~(\ref{XIandPSI})).

\vspace*{0.5cm}

%
\section{Data analysis and results}

Since this work is the first time that 
the Faddeev-Popov procedure  was implemented in NSPT, 
much attention was devoted to reliability checks, which we
describe in the first subsection. The second subsection is devoted to
carrying out the limit in \eq\nr{eq:betaa}. All the numerical values
shown in the following were obtained with $\Nc = 3$.

%
\subsection{Consistency checks}

The first checks were performed against the theory without
gauge fixing, both in $3$ and in $4$ dimensions. In other words, 
we set $\mG \equiv 0$, and checked that we reproduce gauge invariant 
results for the plaquette expectation value, 
irrespective of the gauge parameter $\alpha$ used. 
In particular, we made sure that, for fixed volumes ($V \equiv L^3$), 
we could reproduce the 3d results in Ref.~\cite{PRyork}, 
up to 4-loop order.

\begin{table}[t]
\begin{center}
\begin{tabular}{|c|c|c|}
\hline
$a\mG$ &   exact   & NSPT \\
\hline
0.10   &  2.6579  & 2.6581(6) \\
\hline
0.15   &  2.6481  & 2.6476(6) \\
\hline
0.20   &  2.6379  & 2.6379(8) \\
\hline
0.40   &  2.5713  & 2.5710(8) \\
\hline
0.60   &  2.4714  & 2.4710(8) \\
\hline
0.80   &  2.3485  & 2.3481(7) \\
\hline
1.00   &  2.2119  & 2.2117(6) \\
\hline
1.20   &  2.0689  & 2.0681(5) \\
\hline
1.40   &  1.9251  & 1.9247(5) \\
\hline
1.60   &  1.7848  & 1.7846(4) \\
\hline
\end{tabular}
\vskip 0.5cm
\caption{Comparison between ``exact'' and NSPT results for the 
coefficient of $\beta^{-1}$ for the observable in \eq\nr{observable}.
The lattice extent is $L = 8 a$, except for $am = 0.10$, 
where it is $L = 10 a$.}
\label{T1}
\end{center}
\end{table}

We then plugged the mass terms in. A first test was that the leading 
$\rmO(\beta^{-1})$ contribution to the plaquette expectation value
appearing in \eq\nr{eq:betaa},
\be
 \bigl\langle
  1 - \tilde \Pi_{12}
 \bigr\rangle_\rmi{L}
 = 
 \frac{1}{\beta} \frac{d_A}{3}
 \biggl[ 
  1 - \frac{(am)^2}{N^3}
  \sum_{n_i = 1}^{N}
 \frac{1}{\sum_{i=1}^3 4 \sin^2(\pi n_i/N) + (am)^2}
 \biggr] + \rmO\Bigl( \frac{1}{\beta^2} \Bigr)
 \;, \la{observable}
\ee
where $N \equiv L/a$, 
was correctly reproduced by the NSPT numerics. Table~\ref{T1} 
shows the results for the coefficient of $\beta^{-1}$ 
and gives an idea of the size of errors. 
It also lists all the mass values that we will use in the following.

As a final preparation
we checked, as already explained in Ref.~\cite{lat05},
that if the volume is kept finite, and a fit 
in $(a\mG)^2$ is performed for measurements carried 
out with the masses shown in Table~\ref{T1}, then the 
intercept with the axis $a\mG =0$ agrees with previous results
in the massless theory, obtained without
gauge fixing~\cite{PRyork}.

\subsection{Detailed analysis}

After the consistency checks, we turn to the actual analysis
of \eq\nr{eq:betaa}. It consists of two steps. First, for any given 
value of $\mG$, we need to carry out an infinite-volume extrapolation. 
Second, the infinite-volume extrapolations need in turn to be
extrapolated to $\mG\to 0$, as dictated by \eq\nr{eq:betaa}. 
Both of these extrapolations are rather delicate so let us 
describe the procedure that we adopt in some detail.

\begin{figure}[t]


\centerline{%
\epsfysize=7.0cm\epsfbox{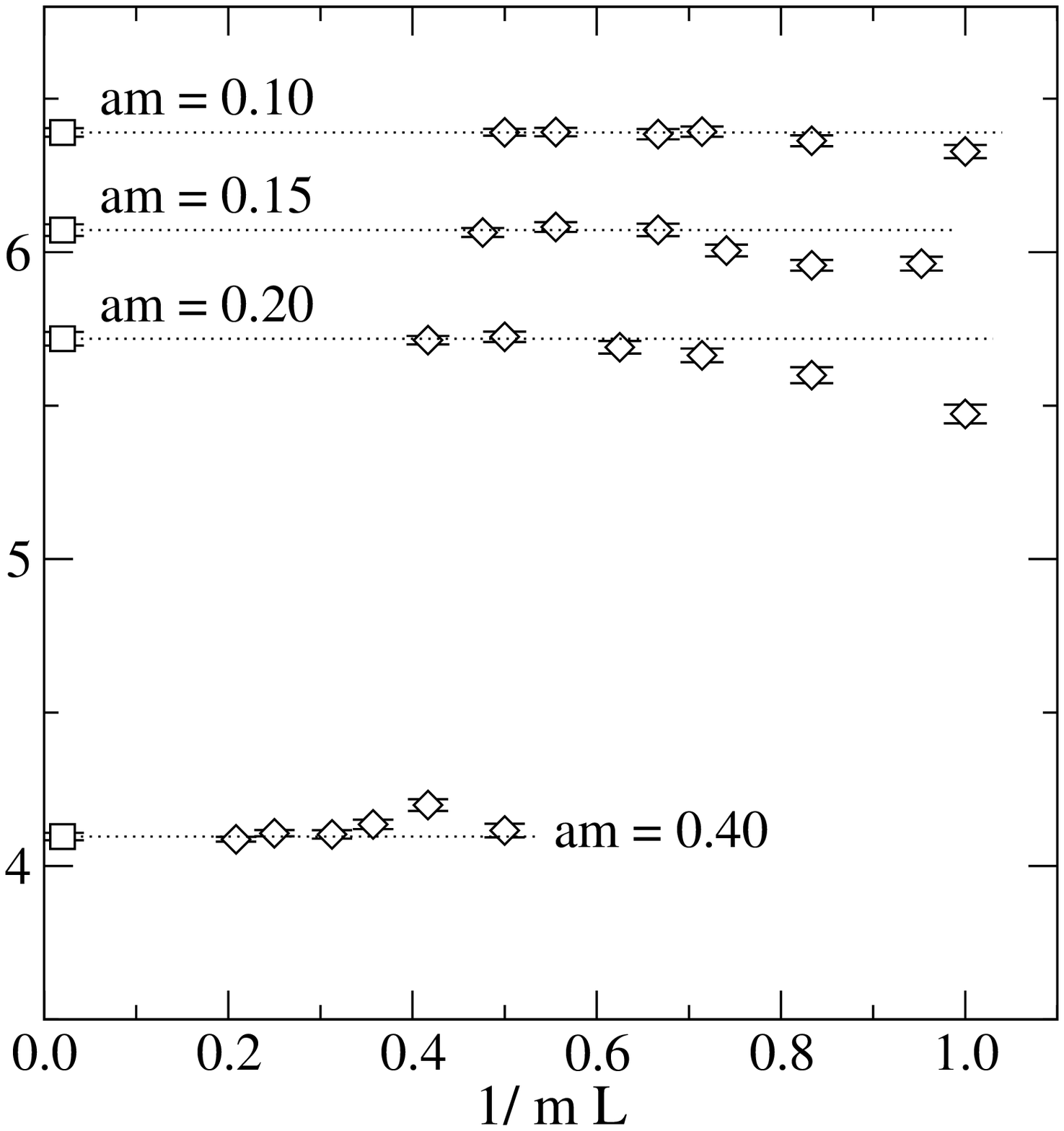}%
\hspace*{1cm}\epsfysize=7.0cm\epsfbox{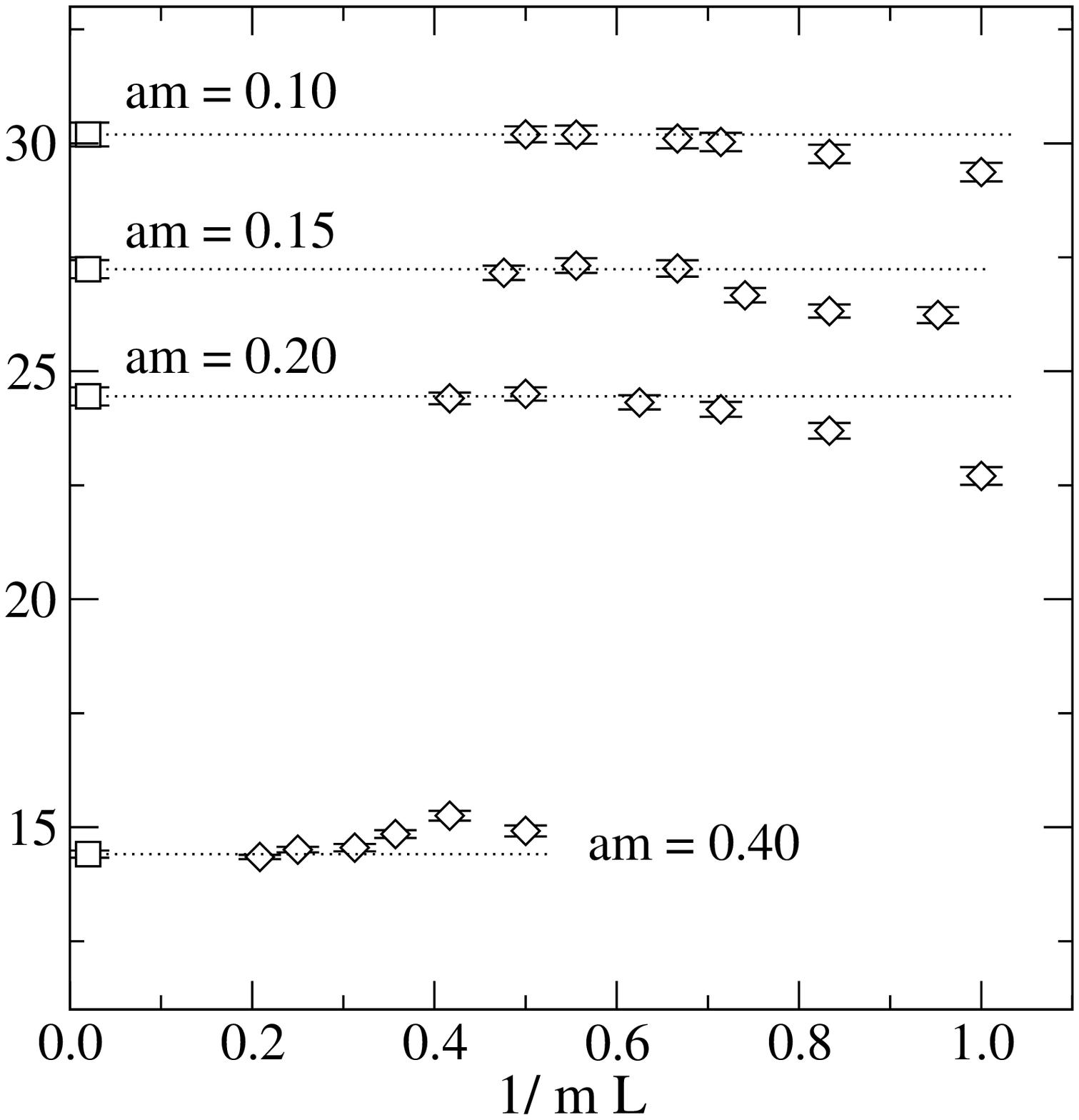}%
}

\caption[a]{Examples of finite-volume values (open diamonds)
and infinite-volume extrapolations (open squares) for the 
coefficient of $\beta^{-3}$ (left) and $\beta^{-4}$ (right) 
in the mass regularized plaquette expectation value, \eq\nr{observable}.} 
\la{fig:mL}

\end{figure}

For the infinite-volume extrapolations, we expect that the dependence
on $L$ is exponentially small, once the volume is large enough. 
Concretely, inspecting the 1-loop expression in \eq\nr{observable}
for $am \ll 1$ but at finite volumes, 
suggests that for $mL \gg 1$ the behaviour is 
\be
 \bigl\langle
  1 - \tilde \Pi_{12}
 \bigr\rangle_\rmi{L}
 \sim \gamma_0 + \frac{1}{\mG L}
 \Bigl[
   \gamma_1 \exp({-\mG L}) +  
   \gamma_2 \exp({-\sqrt{2} \mG L}) +  
   ... 
 \Bigr] 
 \;, \la{voldep}
\ee
where $\gamma_1,\gamma_2,...$ have all the same sign. 
However, for our larger masses $am\gsim 0.40$ (cf.\ Table~\ref{T1}), the 
volume dependence appears in fact to be dominated by 
discretization effects not contained in~\eq\nr{voldep}. 
Moreover, at higher loop orders other structures also 
appear and it is not clear \emph{a priori} how large
$mL$ has to be for them to remain negligible
(note that for the important small masses $am \lsim 0.20$ 
we are only able to go up to $mL \sim 2$, cf.\ \fig\ref{fig:mL}).

For these reasons, we adopt a practical 
procedure in the following whereby we increase the 
volume until no volume dependence is seen within the error bars, 
and then fit a constant to data in this range. To be conservative, 
the resulting error bars are multiplied by a factor two. The original ``raw'' 
data at finite volumes, and the corresponding infinite-volume extrapolations
obtained with the recipe just described, are illustrated in \fig\ref{fig:mL}
for a few masses. However, 
we have also tried other procedures, like a fit according
to \eq\nr{voldep} or to different phenomenological forms ruled 
by decaying exponentials multiplied by polynomial prefactors. 
\fig\ref{fig:mL_2} 
gives an idea of the effects of these variations
on the infinite-volume extrapolations. 
It can be seen that our doubled error bars 
can cover all the variations.

\begin{figure}[t]


\centerline{%
\epsfysize=7.0cm\epsfbox{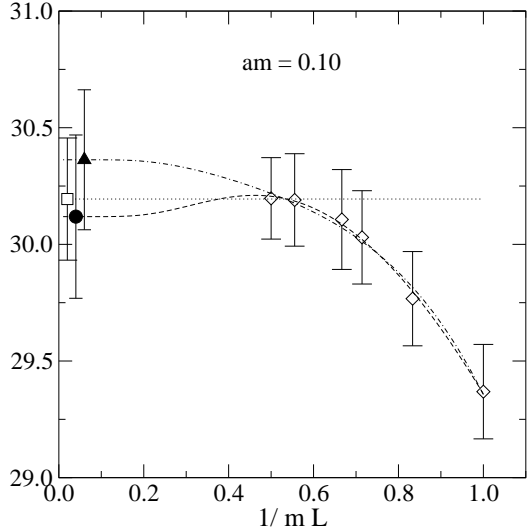}%
}

\caption[a]{An example of different infinite-volume extrapolations 
for the coefficient of $\beta^{-4}$ in the mass regularized plaquette 
expectation value, for $am = 0.10$.
The open square denotes the extrapolation described in the text; 
the closed circle is an extrapolation according to \eq\nr{voldep}
(with unconstrained coefficients);
and the closed triangle assumes yet another phenomenological fit form,  
$\delta_0 + \exp({-\mG L}) [\delta_1 
+ {\delta_2}{(\mG L)^{-4}}]$. } 
\la{fig:mL_2}

\end{figure}

Given the infinite-volume extrapolations, we can carry out
the extrapolation $a \mG\to 0$. Motivated again by a 1-loop
analytic computation (\eq\nr{observable} for $N\to\infty$), 
we use an ansatz allowing for any 
positive powers of $a\mG$. There is the problem, however, 
that the data points are more precise at larger masses: 
the absolute errors decrease roughly as $\sim 1/(am)^2$, {\em i.e.}\
vary by two orders of magnitude. Thus large masses tend to dominate the 
fit, while the most important region should be that of small masses.

\begin{figure}[t]


\centerline{%
\epsfysize=7.0cm\epsfbox{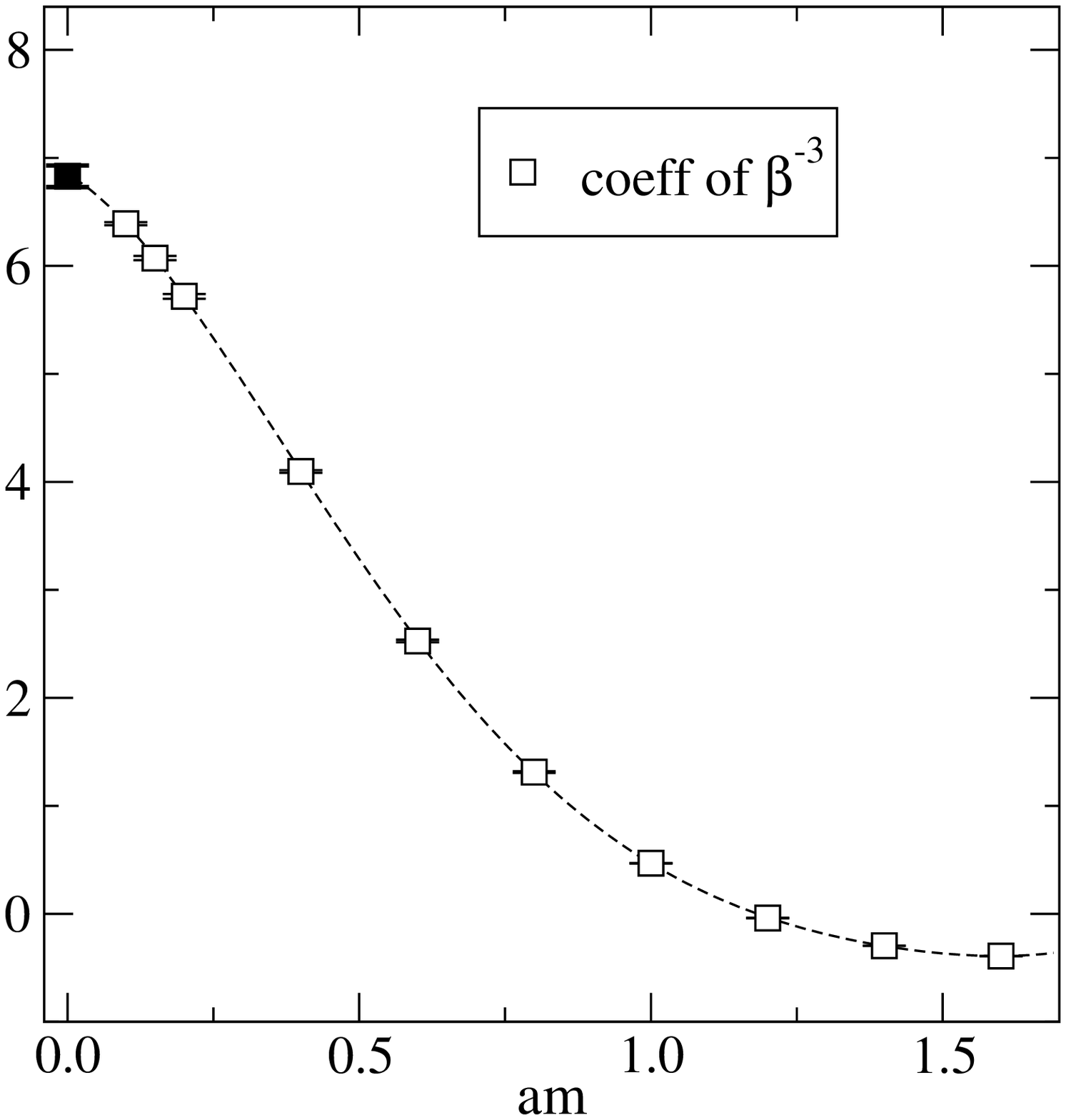}%
\hspace*{1cm}\epsfysize=7.0cm\epsfbox{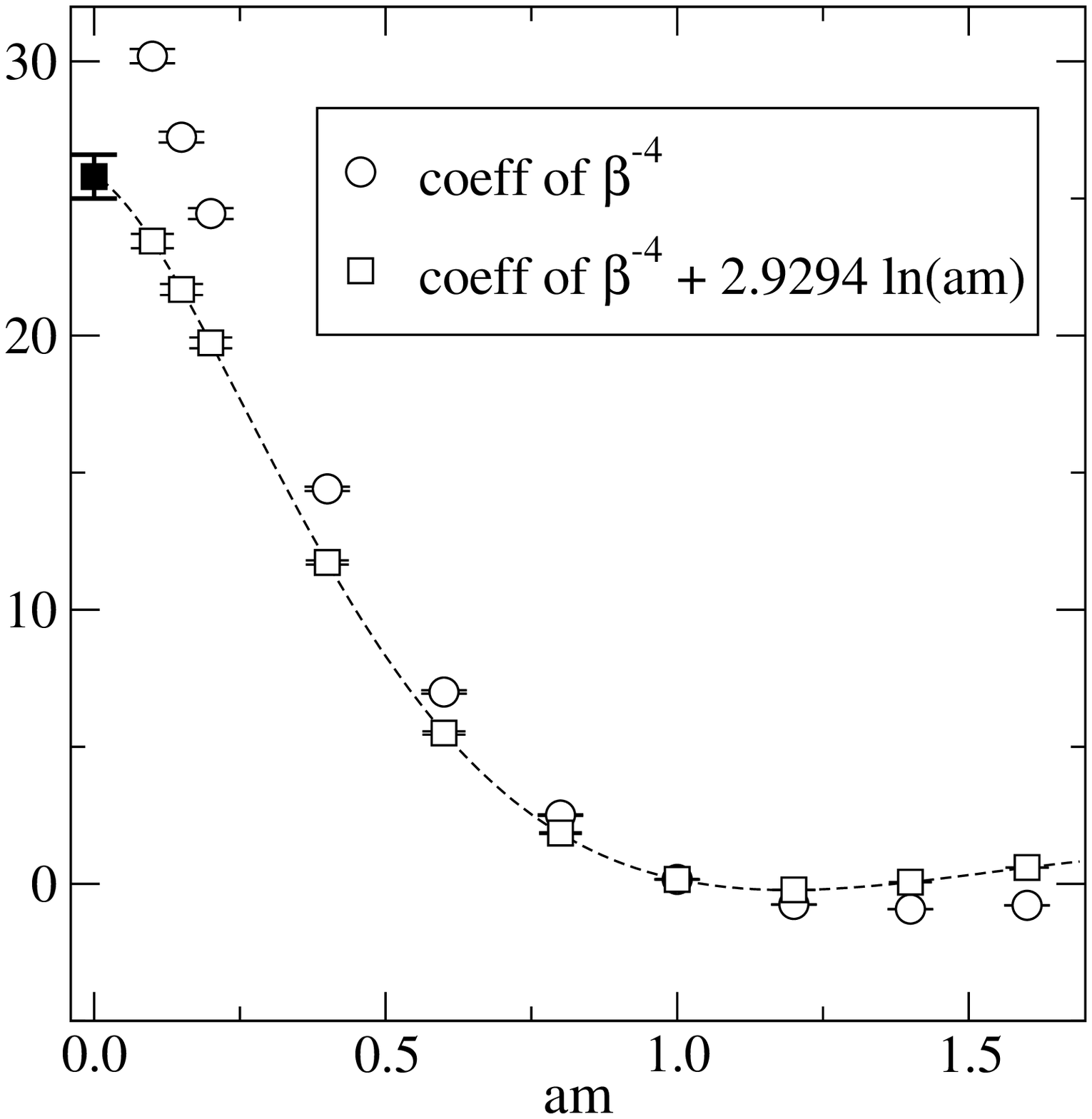}%
}

\caption[a]{The $a\mG\to 0$ extrapolations (dashed lines, with 
the intercepts with the axis $a\mG = 0$ shown with closed squares) 
through the infinite-volume extrapolated data points (open squares), 
for the coefficient of $\beta^{-3}$ (left) and $\beta^{-4}$ (right) 
in the mass regularized plaquette expectation value.
} 
\la{fig:infvol}

\end{figure}

We confront this situation in the following way. 
First of all, we allow for a high-order fit polynomial, 
and monitor the stability of the results, and the $\chi^2$-value
of the fit, with respect to the order of polynomial, as well as 
the number of masses that are taken into account. Second, 
we consider both the regular $\chi^2$-function, and modified 
ones where the errors are weighted by $am$ or by $(am)^2$, 
so as to make the contributions of the different data points 
more balanced. All of these fits generally show an extremum 
as the order of the fit polynomial is increased, with the 
extremal values differing by less than the statistical errors.  
Examples of the fits are shown in \fig\ref{fig:infvol}. 

In order to benchmark this strategy, 
let us first apply it at 1-loop, 2-loop, and 3-loop levels.
We obtain 
\ba
 \lim_{\mG\to 0} \biggl\{ \lim_{L\to\infty}
 \left. \bigl\langle
  1 - \tilde \Pi_{12}
 \bigr\rangle_\rmi{L} \right|_{\beta^{-1}} \biggr\}
 & = &  2.672 \pm 0.008
 \;, \\
 \lim_{\mG\to 0} \biggl\{ \lim_{L\to\infty}
 \left. \bigl\langle
  1 - \tilde \Pi_{12}
 \bigr\rangle_\rmi{L} \right|_{\beta^{-2}} \biggr\}
 & = &  1.955 \pm 0.016
 \;, \\ 
 \lim_{\mG\to 0} \biggl\{ \lim_{L\to\infty}
 \left. \bigl\langle
  1 - \tilde \Pi_{12}
 \bigr\rangle_\rmi{L} \right|_{\beta^{-3}} \biggr\}
 & = & 6.83 \pm 0.10
 \;.  
\ea
These numbers are to be compared with the known results 
in \eqs\nr{eq:c1}--\nr{eq:c3}; we find perfect agreement
within error bars.

We then repeat the same procedure at 4-loop level. 
As shown by \eq\nr{eq:betaa}, 
the extrapolation $a\mG \to 0$ can only be carried out after
the subtraction of the logarithmic IR term. 
The fit shown in \fig\ref{fig:infvol} produces
\be
 \tilde B_\rmi{L}(1)= 
 \biggl( \frac{2\pi^2}{27} \biggr)^2 \times ( 25.8 \pm 0.8 ) 
 = 13.8 \pm 0.4
 \;. \la{tBa}
\ee
Inserting into \eq\nr{master}, a significant cancellation takes place, 
and we obtain 
\be
 \bG = -0.2 \pm 0.4^\rmii{(MC)} \pm 0.4^\rmii{(NSPT)}
 \;, \la{result}
\ee
where the error ``MC'' originates from the lattice Monte Carlo
simulations in Ref.~\cite{plaq}, and the error ``NSPT'' from the 
analysis in the present paper. 
\eq\nr{result} is our final result. 
Let us also record the $\Nc = 3$ values for 
the constant $C_4'$ in \eq\nr{eq:Deltaf} and $c_4'$ in \eq\nr{eq:betaG},  
\be
 C_4' = 10.9 \pm 0.3^\rmii{(NSPT)}
 \;, \quad
 c_4' = 7.0 \pm 0.3^\rmii{(NSPT)}
 \;. \la{Cfinal}
\ee

\vspace*{0.5cm}

%
\section{Conclusions and perspectives}

We have demonstrated in this paper 
the feasibility of determining the non-perturbative 
constant $\bG$, defined through \eq\nr{eq:structure}, by combining
previous lattice Monte Carlo results~\cite{plaq} with a 4-loop perturbative
matching step. The matching involves a comparison of a continuum $\msbar$
computation~\cite{sun} with the corresponding lattice regularized computation. 
The latter we have carried out with the help of Numerical Stochastic 
Perturbation Theory (NSPT). The final estimate for the new matching coefficient
(with two different conventions) is shown in \eq\nr{Cfinal}. 
Taking into account the Monte Carlo results, the final
estimate for $\bG$ is shown in \eq\nr{result}.

We note that 
within the current errors, $\bG$
is consistent with zero. This 
is a matter of conventions, however; for instance, had we not made 
the arbitrary choice of including the factor 2 inside the logarithm 
in \eq\nr{eq:structure}, the corresponding constant would 
be non-zero by a significant amount. 

Given that the physical pressure of hot QCD
is numerically fairly sensitive to $\bG$~\cite{gsixg}, 
it would of course be desirable to improve on the accuracy 
of $\bG$, both on the lattice Monte Carlo and on the NSPT sides. 
For instance, it would be interesting to repeat the current
study with traditional techniques~\cite{pt}.
Moreover, it should in principle be possible to carry out the matching 
leading to \eqs\nr{Cfinal} by using a finite volume rather than 
a mass as an infrared regulator; for this approach the NSPT
side exists already~\cite{PRyork}, but the 4-loop $\msbar$ computations 
of Ref.~\cite{sun} would have to be repeated with 
techniques discussed at 1-loop level for instance in Refs.~\cite{finV}.

Apart from these challenges, there is now an ever more compelling case for 
determining the last remaining purely perturbative $\rmO(g^6)$ term, 
\emph{i.e.}\ the constant denoted by $\bE{1}$ in Ref.~\cite{gsixg}.
Only after this has been added does the $\bmu$-dependence
of \eq\nr{eq:structure} get cancelled, such that the full physical pressure
is scale-independent, as it has to be.

\vspace*{0.5cm}

%
\section*{Acknowledgments}
We thank A. Mantovi for collaboration at preliminary stages of this work, 
K.~Kajantie, C.~Korthals Altes, and K.~Rummukainen for discussions, 
and B.~Kastening for communicating results from Ref.~\cite{bk} 
prior to publication. 
The Parma group acknowledges support from MIUR under 
contract 2004023950\underline{~}002 and 
by I.N.F.N.\ under {\em i.s.~MI11}. F.D.R.\ and Y.S.\ also acknowledge 
support by the {\em Bruno Rossi INFN-MIT exchange program} at an early 
stage of this project. We warmly thank {\em ECT*, Trento,} 
for providing computing time on the {\em BEN} system. 
The total amount of computing time used for this 
project corresponds to about $4 \times 10^{17}$ flop.

\newpage

%

\end{document}